# Parallel paths across the Pacific: a speculative explanation for the *dilep* in Marshallese navigation

Inman Harvey
Evolutionary and Adaptive Systems Group
University of Sussex, Brighton, UK
inmanh@gmail.com

**Abstract**: Traditional techniques used by navigators in the Marshall Islands include the use of wave patterns as influenced by reflection and refraction around islands. The *dilep* is one such pattern, apparently providing signals to guide a navigator directly between two distant islands; so far there is no agreed causal explanation for such a phenomenon. We propose a mechanism; this generates a number of qualitative and quantitative predictions that may in principle be tested against satellite photo evidence, hydrodynamic simulations, experiments by small boat navigators in the right conditions, and ethnographic reports.

## Introduction

Polynesians and Micronesians have in the past successfully navigated vast distances in small boats across the Pacific without the tools that Western navigators have considered essential. It is reasonably easy to appreciate the use of star patterns, and the observation of natural signs such as bird flight to indicate the direction of nearby land. However the use of wave navigation still has some aspects difficult to understand within a conventional scientific framework. This appears to be a highly sophisticated practice, learnt through extended training at detecting the interactions of multiple swells merging with each other, reflecting from and refracting around islands. One wave navigation mystery is the *dilep*, translated as 'backbone' or 'spine', that apparently forms a wave path for sailor-navigators in the Marshall islands to follow, leading directly between islands more than 100km apart. Genz et al (2009) and Huth (2016) have lengthy discussions of *dilep*, and the other referenced publications by the same authors provide an invaluable context. Though navigators have demonstrated the ability to detect such a wave path, through the specific pitch and roll patterns of their outrigger canoes, there is till now no agreed scientific explanation for this.

In this paper we shall ignore most of the practical problems, and adopt a brutally simplified scenario to model, to see if there is in principle any form of invariant signal. We initially assume just two point-islands in an ocean otherwise featureless bar a reliably constant primary swell from a fixed direction. The islands each reflect secondary swells of the same or similar period, and we seek potential signals in the interactions between these secondary swells. Since reflected waves have much less energy than the primary swell, e.g. at best 10% (Huth 2016), and this decreases as they expand outwards, the best opportunity for a discernible signal is where the two reflected swells act additively. We plot the regions where in principle this happens maximally, which turns out to be a matter of plotting elongated ellipses with the atolls at the foci. We then explore how this idealised model may change on the introduction of lesser synchronisation and limited noise.

From this core idea we produce a number of results, some potentially verifiable by experiment, that we shall list here before elaborating in the body of the paper.

1. We plot such multiple elongated elliptical pathways of standing waves between two islands, that we wish to identify with *dileps*. (See e.g. Figure 3 below)

> This proposal for multiple (near-)parallel *dileps* is the most radical and challengeable discrepancy with the conventional view of just one *dilep*. All other consequences flow from this.

2. Our model predicts that useful signals will only be available to a navigator who is currently on a *dilep*, and these signals relate to boat-headings and not (as antecedent models apparently suggest) to boat-positions relative to a *dilep*.

3. We provide a quantifiable prediction of the expected distance between the 2 most central parallel *dileps*. For two islands N wavelengths (of the primary swell) apart, this expected distance is around 0.7-1.0 times sqrt(N) wavelengths. In the scenario outlined here this is around 2 or 3kms. Distances further out are also predictable.

4. We discuss variants where the reflected swells do not have identical wavelengths, where the wavelengths and periods intermittently change, or where there are multiple reflections off irregularly shaped islands or atolls.

5. We provide various alternative characterisations, at different scales, that could potentially fit the Marshallese concept of *booj*, distinct entities that apparently are perceived in sequence whilst following a *dilep*.

6. We briefly discuss possible cues available to the navigator using a *dilep*.

7. We propose a possible resolution of a perceived contradiction between different Marshallese conceptual schemes. One view (as used here) is that *dileps* arise from secondary waves reflected from the islands. The other view is that *dileps* arise from opposing swells at right-angles to the path. This second view has been difficult to understand, but may plausibly conceptualise the patterns of many (near-)parallel standing waves that our model shows.

Our model is highly speculative, with only very indirect contact with data from Marshallese waters. Once the big gamble has been taken on just how much can be thrown away in the minimalist model, the rest follows inevitably. No attempt has been made to assess whether such signals are sufficiently strong to be in practice discernible. The argument takes the form of asserting that, however improbable, these patterns appear to be the only possible invariant candidates available for *dileps*; we then generate testable predictions from this hypothesis. If this is wrong, it is hoped that the large number of operationally defined predictions should make it relatively straightforward to disconfirm.

**Antecedent Models**

The literature (e.g. Genz et al, 2009; Huth, 2016) presents a basic picture of how *dileps* are experienced and used by traditional navigators, based largely on their verbal reports together with some limited observations of them on a few voyages. A typical such model is presented in Figure 1: the *dilep* is thought of as a single track, leading directly (or nearly directly — sometimes it is reported as not completely straight) from starting island A to destination island B, against a background of some primary swell that may be transverse to the required route.

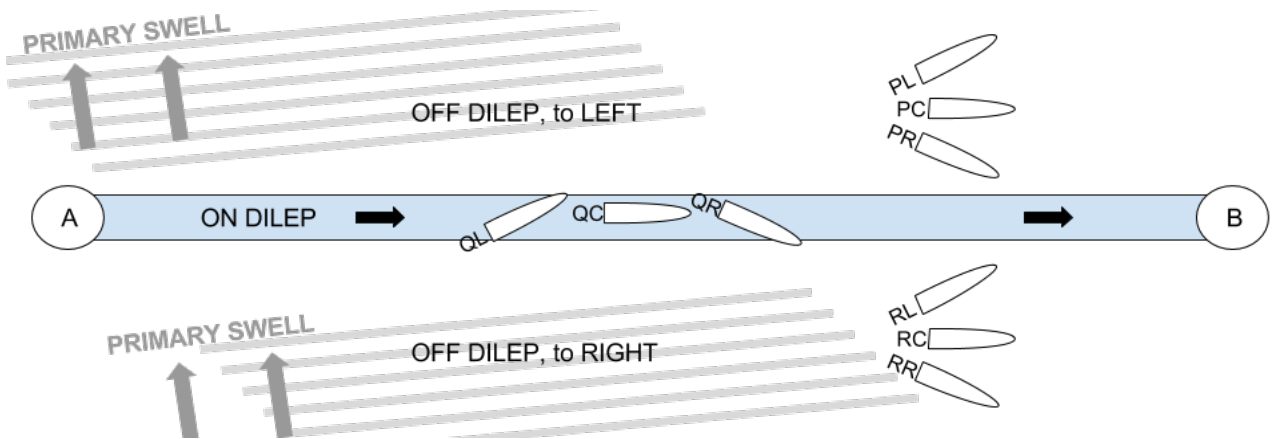

Figure 1. A sketch of typical features of antecedent *dilep* models, against a background of the (only partially shown) primary swell. The example boats differ in *dilep*-relative positions (prefixes P, Q, R) and in headings (suffixes L, C, R).

The primary swell, probably long and slow, provides a generally reliable directional frame of reference. The *dilep* apparently does more than this, providing a unique track targeted at the destination island, with signals perceivable through the motion of the boat. The mechanism generating such signals is a mystery in these antecedent models, and we are given little guidance as to the width of this track, and whether its edges are sharply defined or fade away gradually.

We should distinguish between a track and a heading. In Figure 1 boats marked QL, QC, QR are all located on the same *dilep* track, but with different headings: respectively left-of-course, on-course, and right-of-course to the destination. The signals described in these antecedent models apparently indicate track-relative positions and fail to discriminate between headings:

> According to Captain Korent, while on the *dilep*, the vessel has a symmetric rolling motion; if the boat strays off the *dilep*, the vessel acquires a more asymmetric motion.
> Huth, 2016, page 165.

It would seem, from this perspective, that vessels QL, QC and QR would all share a similar symmetric rolling motion. PL, PC and PR presumably all share the asymmetric motion associated with 'left of track'; RL, RC, RR the motion associated with 'right of track'. Such signals would only change when crossing the edge of the *dilep* track, and would not change when e.g. boat QL changes its heading to that of QR.

The model we develop here has some similarities with, but then some very radical differences from, such antecedent models. We agree on the pragmatics of a navigator distinguishing between three signals L, C and R, and thus turning appropriately. But in our model these signals refer to headings rather than track-relative positions, and are only available actually on the *dilep* as QL, QC, QR. Off the *dilep* — between the multiple near-parallel *dileps* of our model— we expect *dilep*-related signals to fade away, and only the primary swell to be available to give any sense of direction.

## Our Basic Model

We assume an ocean has a constant primary swell from some fixed direction. For visualisation purposes we may assume that it is a tropical swell of period 8 secs, wavelength 100m, conveniently round numbers; this wavelength becomes the primary unit of length in the model.. Suppose there are two atolls A and B, conceptualised as points, distance N = 1000 wavelengths apart (i.e. 100km).

These each reflect the primary swell with, we shall initially assume, the same period and wavelength. We simplify the model by now removing the primary swell altogether, just treating A and B as two point sources of identical secondary swells, expanding radially. How do they interact?

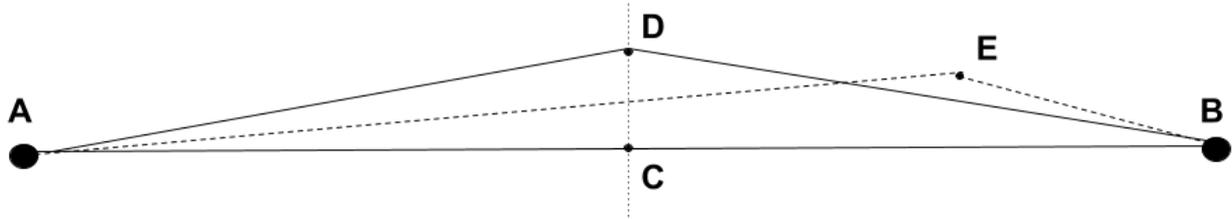

Figure 2. C is midway between atolls A, B. Swells of equal periods emanate in phase from A, B. If AD+DB is an integral number (*alternatively*: +0.5) of wavelengths, they maximally combine additively (*alternatively*: destructively) at D. The same is true for any E, such that AE+EB=AD+DB. This defines an ellipse with foci A,B, and semi-minor-axis CD.

If they are exactly in phase, and AB is exactly 1000.0 wavelengths, the two swells will (in principle) combine additively to form a standing wave, like a plucked guitar string, along the direct line AB, (Figure 2). We may identify this as a *dilep*. At other places on the ocean surface, the two swells will cancel each other out. For example consider point D, offset from mid-point C by an amount that makes AD=DB=500.25 wavelengths. Here, and for any other point E such that AE+EB=1000.5 wavelengths, the waves combine destructively. This describes a very narrow elongated ellipse, with foci at A and B, that passes through D; we can call this ellipse where wave activity cancels out an anti-*dilep*. If we take a further-out ellipse, such that now AE+EB=1001, once again the waves act maximally additively, which we take as being a further *dilep*.

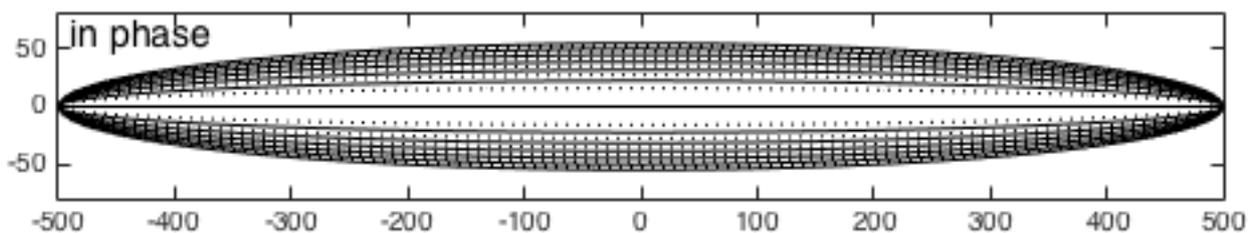

Figure 3. Solid lines denote *dileps*, dotted anti-*dileps*, on the assumption that A, B swells are exactly in phase and AB=1000.0 wavelengths. Only the 11 most central *dileps* are plotted here.

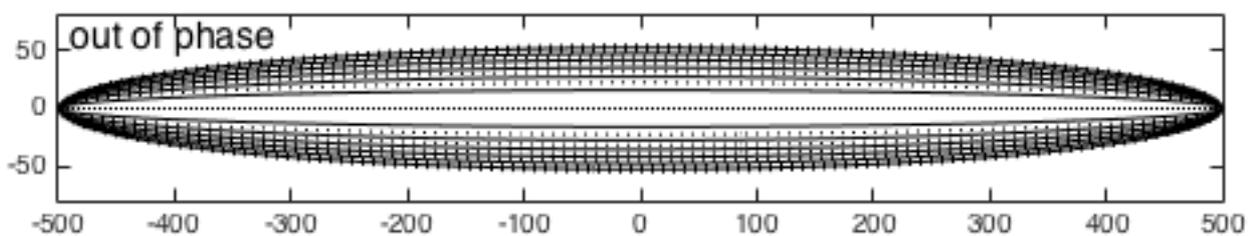

Figure 4. As Figure 3, but now assuming swells are exactly out of phase at A, B. *Dileps* and anti-*dileps* have swapped positions.

This process can be continued, and in Figure 3 we plot a number of such *dilep*s and anti-*dileps*. Figure 4 shows the equivalent when A-swell and B-swell are exactly out of phase at their atoll origins. Reflections from each atoll are likely closest in strength, and hence likely to maximise the prominence of *dilep*s, midway between the atolls where they are of most use to a navigator. Figures 5 and 6 shows this region for the in-phase and out-of-phase cases. We note that the *dileps* get closer together as one gets further away from the centreline; we primarily consider the fairly central ones.

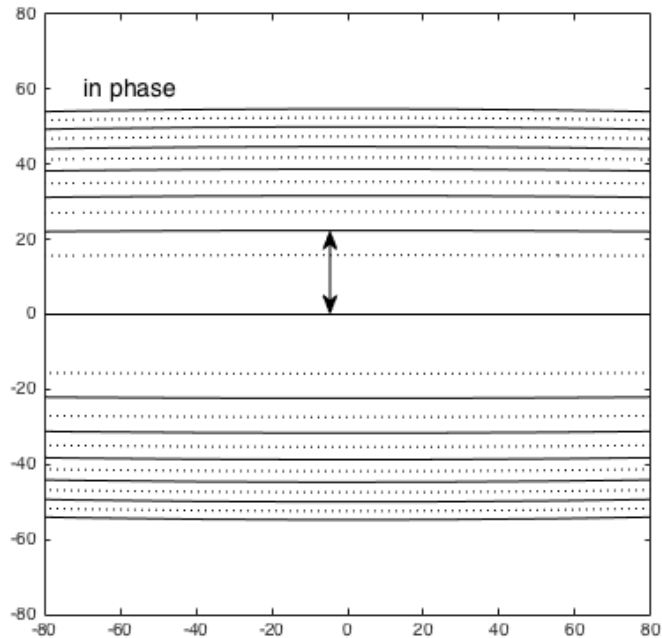

Figure 5. Expanded view of central part of Figure 3, in phase, only 11 most central *dileps*. Arrow denotes distance between central *dilep* and next out out, here around 22 wavelengths, or 2.2km.

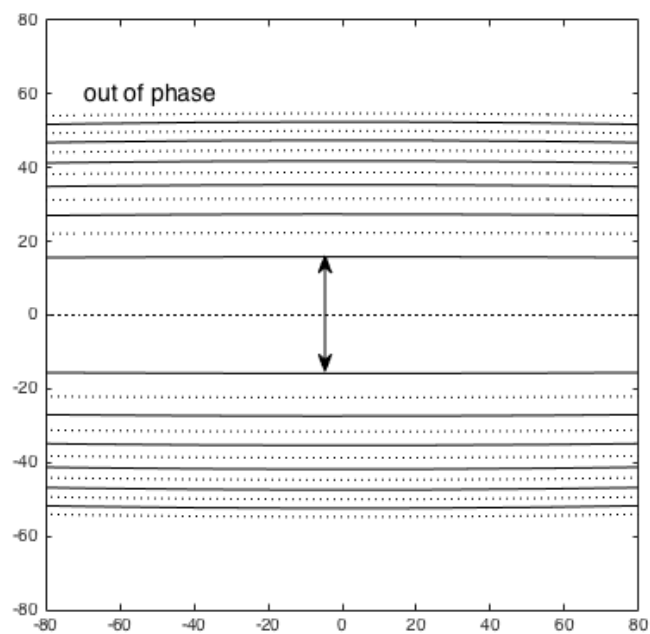

Figure 6. Expanded view of Figure 4, the out of phase case. Arrow shows maximum distance between *dileps* is now 3.2km.

## How far apart are *dileps*?

From Figures 2 and 5, this specific in-phase example gives the distance between central *dileps* as CD when AC=N/2 and AD=N/2+0.5. From Pythagoras' Theorem this is sqrt(N/2), here 22.3 wavelengths or 2.2km. But this may not be typical, so we consider variants.

The 100% out-of-phase example of Figure 6 indicates the gap as 2*CD, where now AD=N/2+0.25. Again from Pythagoras, CD=sqrt(N/4), hence the gap is sqrt(N), here 31.6 wavelengths or 3.2km. If the phase difference is slowly shifted through intermediate values, the ellipses will shift to intermediate regions; comparable changes would result from N, the distance AB in wavelengths, being no longer an integer. The central straight-line *dilep* will either (depending on direction of phase-shift) split open into an initially very narrow ellipse or (if shifting the other way) collapse in on itself and disappear. The lines shown merely illustrate the peaks of the standing waves, with shoulder values either side. Hence it becomes a judgment call, dependent on what threshold values are deemed appropriate, when to identify the moment one *dilep* effectively divides into two, or effectively disappears. Nevertheless we suggest that over a full range of such variants, the largest gap potentially visible between neighbouring parallel *dileps* will lie in (or close to) the range sqrt(N/2) to sqrt(N) wavelengths. Since this may be of the order of 2 to 3kms for the scenario used here, our hypothesis provides two testable predictions; the first that there will be several parallel patterns, and the second quantifying the largest gap between them. Potentially this could be checked by satellite photos, or by traversing across with a local experienced navigator able to discern them.

The plotted lines represent just peak activations for supposed *dileps*, and their effective breadths will scale with gaps between them. Narrower *dileps* occur further away from the direct AB line; for illustration, the 10th and 11th *dilep*-ellipses, counting out from C, will be roughly 7.09km and 7.44km from the centreline at their furthest, and hence around 350m apart..

Figure 7 shows simulation results from islands around 80 wavelengths apart; the ellipses are much fatter than for 1000 wavelengths. Within the central ellipse a somewhat shaded central line can be seen. If this were judged well-enough developed to count as a *dilep*, this implies a central inter-*dilep* gap around 5 wavelengths (from centre-line to first ellipse); if judged otherwise, around 10 wavelengths (minor axis of first ellipse). In either case this supports our rough estimate based on sqrt(80) ≅ 9. The phase-shift in between each *dilep* shows clearly in this image.

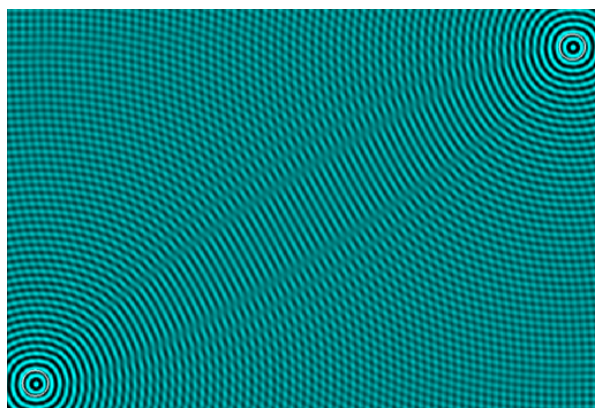

Figure 7. Simulation from rippletank app for smartphone, www.falstad.com, by Paul Falstad; ellipses seen between sources around 80 wavelengths apart. Such simulations need fine temporal resolution to avoid the ellipses appearing to shift, a form of temporal aliasing; the 'wagonwheel effect' seen in old films where wheels may appear to go backwards.

## One *dilep* versus several *dileps*?

The literature on *dileps* apparently assumes there will be just one *dilep* between a pair of islands, whereas this model assumes several. How can this be reconciled?

Although we hope that such multiple parallel patterns could be identified visually from satellites, this is less likely to be possible from a single boat actually using them -- and certainly not if they are kms apart or even 350m. A navigator using *dileps* to navigate from A to B will, having identified one, attempt to track it for as long as possible. He may lose track of it for a variety of reasons: loss of attention, or because (as we shall suggest below) such a pathway may be intermittent, or just through bad luck. If, after searching for a time, he finds the crucial signals again, why would (indeed, how could) he identify it as a different *dilep*? If it feels similar, points in the same direction, serves the same vital goal, then for all intents and purposes it *is* the same *dilep*.

We note that our proposed multiple *dileps* would be much more useful to a navigator than a single central one. This is still compatible with the practical users of such patterns assuming and acting as if there is just one. This could form an interesting case study for cognitive theorists interested in the ontology and epistemology of invariants and/or objects, particularly with a practice-oriented approach to cognition and perception for a situated and (boat-)embodied participant navigator.

## Relaxing exact synchrony

Our idealised model has so far assumed the two reflected secondary swells share identical wavelengths, inherited from the primary swell; and there are no currents. Even the smallest difference in wavelength makes a difference, as illustrated in Figure 8. Here we give 3 examples where the B-swell retains a 100m wavelength, but the A-swell is reduced to 99.9, 99.8 or 99.7m; equivalently, the inter-atoll distance of 1000 B-wavelengths corresponds to respectively 1001, 1002, or 1003 A-wavelengths. We see the *dilep*-ellipses are slightly distorted, retain one focus at/near A, but start to terminate before reaching B. These terminations we shall call *dilep*-ends, and note that where the difference in number of wavelengths (A-swell versus B-swell) between A and B is *n*, *dilep*-ends carve up AB into *n* segments.

These plots are so far based on just wavelength changing, but period remaining the same. In fact we may expect a change in period to be associated with the change of wavelength. Given that speed = wavelength/period, and the accepted rule that speed (in m/s) = 1.25*sqrt(wavelength in m), we have for the original swell: wavelength 100m, period 8s, speed 12.5 m/s = 45.0kph; whereas for a modified swell, e.g that of wavelength 99.7m, we anticipate a period 7.988s, speed 12.481 m/s = 44.93kph. Taking such minor speed differences (here 0.07kph) into account, the effect is to maintain nearly the same overall picture of Figure 8 with each *dilep*-end moving to the right, at a rate equalling half the difference in speeds for each swell. Each ellipse expands and eventually takes over the current position of next-in-line. New ellipses emerge rightwards from atoll A, and the snapshots of Figure 8 catch the moment before a new such mini-ellipse is about to emerge.

Extrapolating to a larger difference of 1% in wavelengths -- A-swell 99m versus B-swell 100m -- we would have *dilep*-ends dividing AB into 10 segments of 10km each. In this case each *dilep*-end would be shifting to the right at 0.113kph, taking some 88 hours for the slow ellipse-shift to complete a full cycle. This slow shift would probably not be discernible to a navigator, and have little practical consequence.

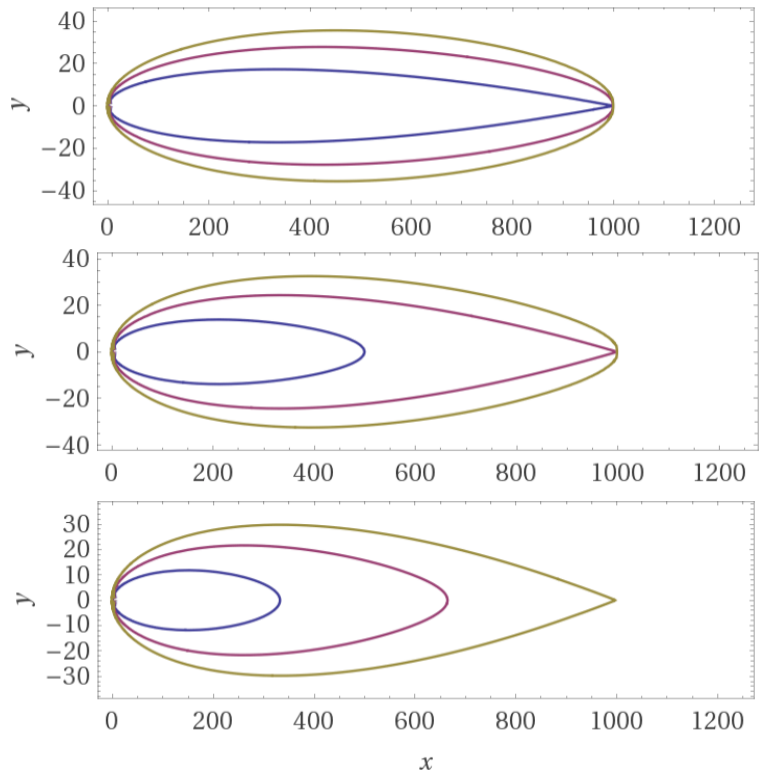

Figure 8. How central *dileps* shift when wavelength of A is reduced to: Top 99.9m, Centre 99.8m, Bottom 99.7m. B wavelength stays at 100m. Wolfram Alpha plot.

In such an idealised world, where the A-swell is systematically slightly shorter than the B-swell, we note that a navigator able to reliably follow such *dileps* would have different experiences depending on direction of travel. Going from B to A, the first *dilep* met can be followed all the way to the destination (deviating somewhat from the most direct straight-line route). But going from A to B (on or near the most direct route) will at best involve repeatedly losing track of a *dilep* when it peters out at a *dilep*-end (here, every 10km), and then needing to search for the next one.

There are many possible reasons why periods should diverge somewhat from a common standard, They change in any case slightly over time and distance, varying currents have an effect, the reflecting atolls are irregular in shape rather than point sources. When a wavetrain includes e.g. subtrains with variant periods, these will tend to travel at different speeds and separate out into spatially distinct extended groups. In all these ways, and more, the real world will diverge from our idealised model. We anticipate that there may well be multiple intermediate *dilep* segments, some short. Nevertheless it appears that whenever there are swells of fairly similar periods from A towards B, and from B towards A, there will be invariants at some scale relating to our ellipses, with the predominant directions of what we want to identify as *dileps* all pointing locally in the direction that suits the navigator's needs.

**Relaxing the idealised perfection of reflections**

Our idealised model has so far assumed the two reflected secondary swells can each be treated as a new perfect point source. In practice an island or atoll is likely to have an irregular outline, and the reflection from one prominent headland will be displaced by many wavelengths relative to that from another headland. Some such reflections will interact constructively, some destructively; overall, will that generate extra interference that jeopardises the effectiveness of the reflections as *dileps*?

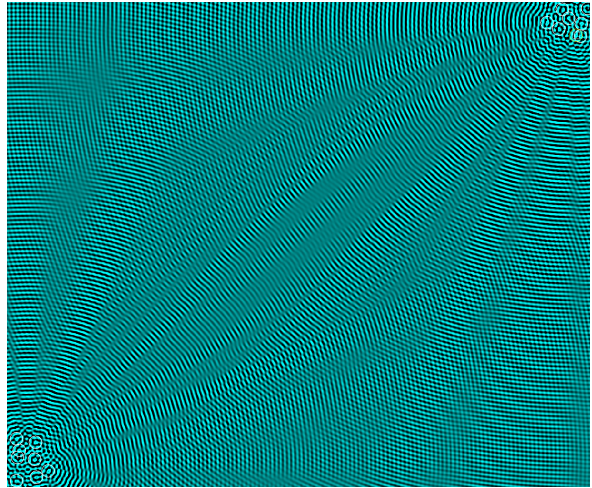

Figure 9. As Figure 7, but here irregular islands at top right and bottom left are each represented by a cluster of 8 point sources, haphazardly positioned.

We cannot yet answer this definitively. But we can show one simulated example in Figure 9. Each island is here modelled as 8 different point sources haphazardly placed in a cluster. Comparing this with the earlier Figure 7 that used single point sources for each island, we note that new constructive patterns of interference are visible. But we also note that there are still visible remains of some potentially useful *dileps* in the inter-island region. Whilst inconclusive, this is at least hopeful.

**What is a *booj*?**

We note from the literature (Huth 2016, Genz et al 2009) the Marshallese concept of *booj* variously translated as a 'knot' or a 'node of intersection of swells'. Genz et al's Figure 5 clearly show these as distinct entities to be encountered along a *dilep* pathway, but we are not sure at what scale they might appear. Based on our model, we have various candidates at very different scales.

At the shortest scale we note that, along the *dileps*, the reflected waves from each atoll are in sync and add up to form a standing wave as on a plucked guitar string. The distance along any one *dilep* between nodes of maximum activity will be half a wavelength, i.e. here 50m, so we would expect a boat to experience peaks of pitch-activation at these intervals, which in the scenario outlined here means around every 50 to 25 seconds for a boat travelling at 1 to 2 m/s (around 2 to 4 knots). If indeed such peaks can be reliably detected, they would provide an excellent opportunity for calculating speed -- they literally measure distance covered over the sea-floor. Between peaks of pitch-activation (at 50m intervals) there are also peaks of heave-activation (similarly at 50m intervals), but we shall suggest below that the pitch is likely to be more discernible than the heave.

As an intermediate range phenomenon, there are many reasons why variations in sea conditions, such as one wavetrain with a specific wavelength followed by another with a slightly different wavelength, might cause conditions to vary over periods of perhaps several minutes, Each such change could occasion the loss of one *dilep*, and hopefully the re-creation of a different one that is (near-)parallel and  not too far away. For illustration, a wavetrain of 20 waves, wavelength 100m, individual wavespeed 45kph, has a group-speed of half that in deep water, and presumably takes around 5 or 6 minutes to pass a point; such typical underlying statistics of sea-patterns would translate into the statistics of *dilep* intervals. We have already outlined above why the navigator seeking to use such *dileps* to navigate from A to B will have no reason to consider the new one a *different dilep*. But this intermittency might explain why different segments -- *boojs*? -- could be experienced over a scale of several minutes.

At a longer timescale yet, perhaps several hours, we noted above (eg Figure 8) that when the A-swell is systematically slightly shorter wavelength than the B-swell, navigating by *dilep* from A to B involves repeatedly encountering (and then losing) *dilep*-ends. With the example discussed of 1% shorter wavelength, intervals of 10km, at boat speeds of around 1 to 2 m/s these would be met say every 1 to 3 hours. The plots show a sharp line for peak activation of the standing wave, but when taking into account the shoulder values, a *dilep*-end represents an area of near-peak activity that will be greatly extended in both width and length. We may conjecture that here the *dilep* loses its capacity to point towards the destination and the navigator would need to revert to maintaining course in relation to the primary swell. This is the last of our 3 possible candidates for a *booj*: candidates to be encountered on timescales of seconds or minutes or hours.

**Marshallese schools of thought on *dileps***

The literature by Genz and Huth talks of two different schools of thought used in the Marshalls for explaining *dileps*. On the one hand one school talks of reflected waves in a way that matches the assumptions made in this paper. On the other hand, one of the last traditionally trained Marshallese navigators, Captain Korent Joel (who died in 2017; a co-author of Genz et al 2009), talked of the underlying cause being opposing swells coming from each side of the inter-island path, and a further informant Thomas Bokin agreed with him.

> The explanation of Bokin and Joel that it is created by intersecting and opposing swells doesn't work terribly well for our usual Western understanding. Their description implies that two opposing swells are somehow always oriented parallel to the path connecting pairs of islands. (Huth, 2013)

We have argued that the navigator actually using a *dilep* to navigate in practice between A and B has no reason to think there is more than a single *dilep* pointing the way. But a Marshallese theoretician may seek a bigger picture and might through experience have developed a version of our model with multiple parallel *dileps*. For example, a transit across at right-angles to the AB path could reveal a succession of broad standing waves as each *dilep* was crossed. And this would look remarkably similar to the succession of standing waves produced by two opposing swells. This second explanation fairly explicitly proposes multiple parallel *dileps*, not just one.

In Figure 10 we show how the big picture varies between the two Marshallese schools of thought, but then focus down on a smaller area of ocean, the 2.4km square box indicated, to see (in Figures 11, 12) how the two types of explanation compare at such scales.

Figure 11 shows the peak lines of standing waves based on the ideas presented here: secondary swells of period 8s, wavelength 100m, radiating from each atoll are here roughly 350m apart — with the gaps decreasing slowly as one moves further from the centreline. Figure 12 shows the standing waves generated by the competing explanation. If we assume the notional swell has wavelength 700m, the standing waves would be consistently 350m apart.

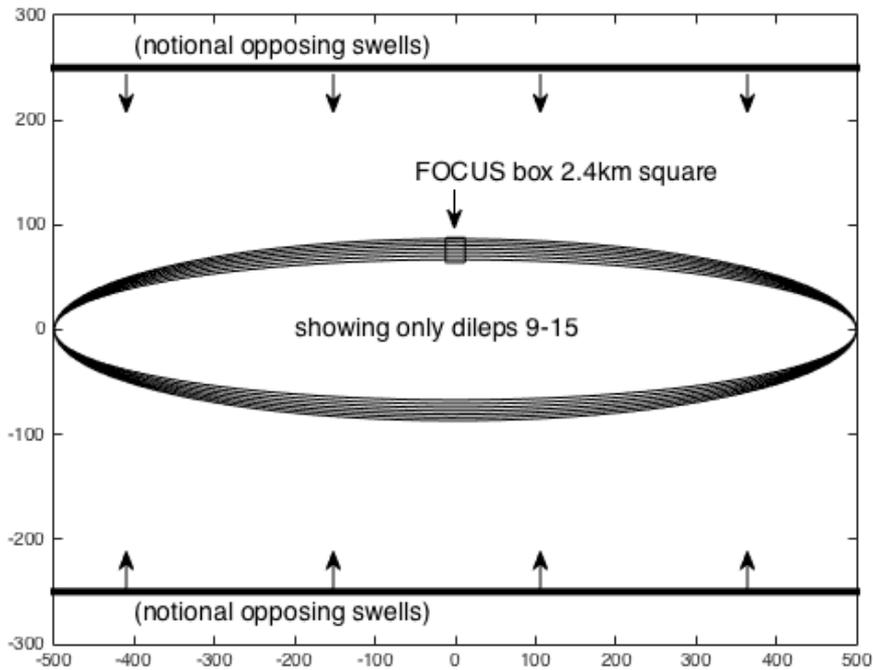

Figure 10. The ellipses show the *dileps* generated from swells radiating from the atolls L&R, corresponding to the first Marshallese school of thought (only *dileps* 9-15 plotted here). The second Marshallese school posits notional opposing swells as illustrated top and bottom. We focus on the 2.4km square area shown to see how the theories compare.

The navigator relying on the second explanation might experience a dilemma if asked to estimate both wavelength and period of the notional opposing swells. If any distance cues are discernible, they would support a long wavelength, here 700m. However temporal cues, from the oscillations of the standing wave, support the period of 8 seconds. Usual wavelength::period relationships suggest 700m::21s and 100m::8s. Despite such possible quantitative mismatches, the striking qualitative similarity between Figures 11 and 12 makes the second conceptual framework very plausible, even if the first is the correct one; this further supports the idea of many parallel *dileps*, not just one.

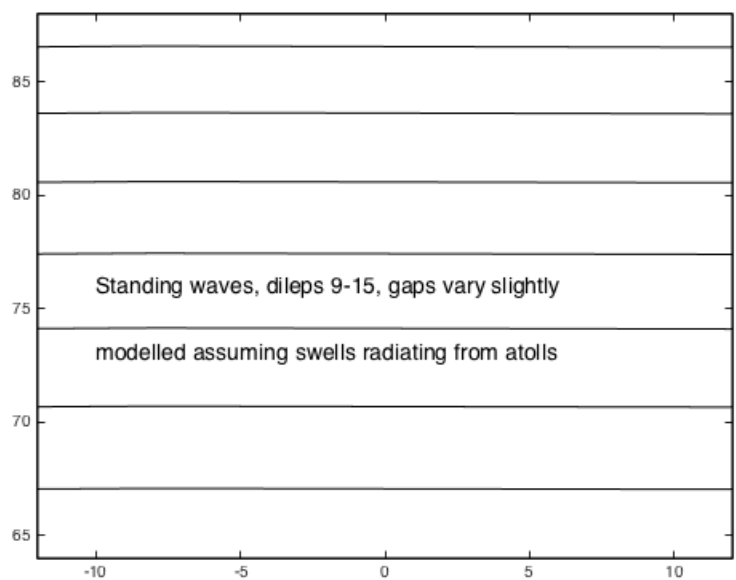

Figure 11. Lines of standing waves, i.e. *dileps*, according to the model presented here based on swells radiating from the atolls. Note that the gaps decrease slightly towards the top.

Genz et al (2009) discuss a journey of 100km travelling north from Majuro to Aur in a primary eastern swell:

> Several times during this northern voyage, Captain Korent described a rocking motion from intersections of east and west swells. He followed a succession of these *booj* along the *dilep* toward land, but the wave buoy data from these locations indicated only an east swell.

This is compatible with our assumptions that the only *real* relevant swells are here one primary eastern plus two weaker reflected radiating secondary swells (that in mid-journey are roughly northern and southern). The two further 'east and west swells' that Captain Korent describes are, we suggest, conceptual only; but nevertheless provide a largely plausible explanatory framework for the wave-patterns experienced, particularly were there to be many parallel patterns.

**Cues for the navigator**

This paper focusses on where, in an idealised world, a signal might exist. For the most part we leave aside practical issues as to how such signals might be detected, though here we briefly consider some of the possibilities.

We use a boat-centred frame of reference, based on a fore-aft axis, a port-starboard axis, and a vertical axis. Rotational motions about such axes are termed respectively roll, pitch, and yaw; linear motions along such axes are respectively surge, sway and heave. All such motions, from primary, secondary, tertiary swells and beyond, will be combined in the dance of the boat on the ocean. As well as the size of such motions, the periodicity and synchronicity can be crucial.

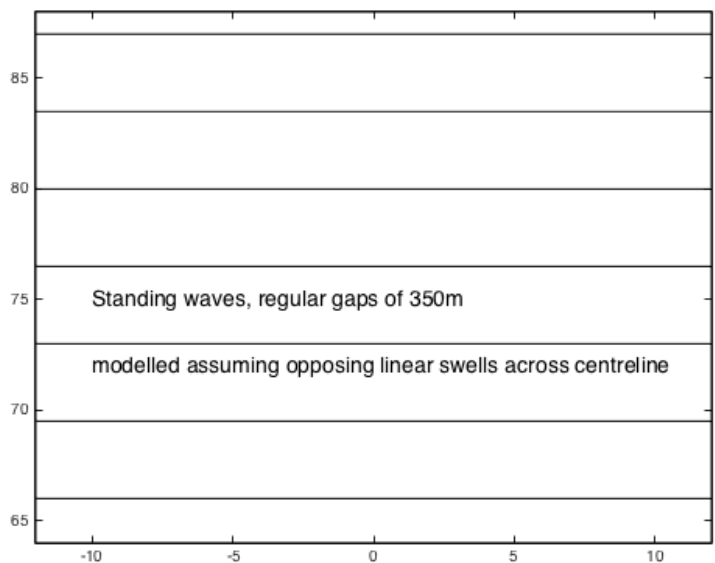

Figure 12. The alternative explanation, based on notional opposing swells of wavelength 700m, would generate standing waves as shown, regularly 350m apart.

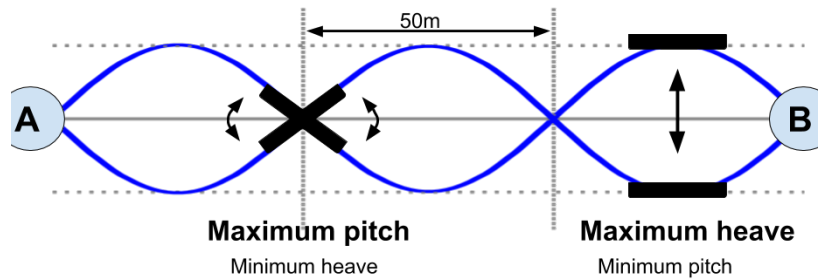

Figure 13. Heave and pitch from the reflected secondary swells, as experienced by a boat travelling along a *dilep*; just 2 example boat positions are shown. The primary swell (and any others) will add their own effects.

As illustration, Figure 13 is an idealised version of how a boat travelling on the *dilep* standing wave we propose will move between regions of maximum pitch + minimum-heave and regions of minimum pitch + maximum heave. The heave signal from these secondary reflected swells, with a period of 8 seconds, looks significant here, but it may well be lost when one adds in a substantially larger heave effect, with the same period, from the primary swell.

The maximum pitch cue, however, may be much less affected by the primary swell. If the latter is at right-angles to the secondary, it makes no contribution to the pitch; if at an oblique angle, the primary will only partially affect pitch, simultaneously with its larger effect on roll and heave.

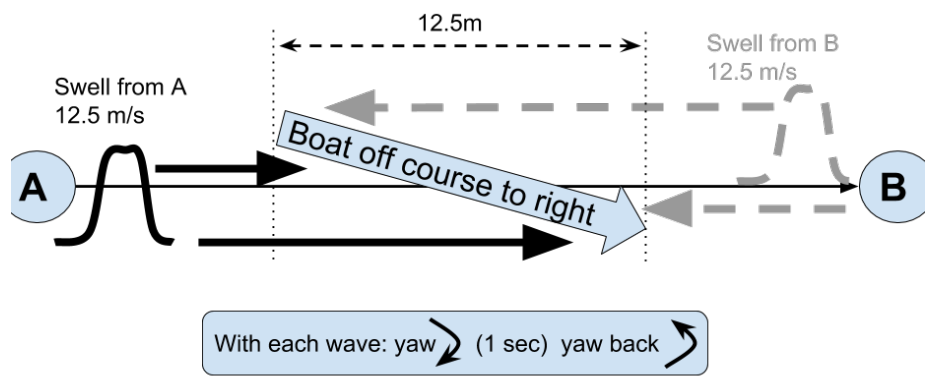

Figure 14. A 12.5m long boat heading for B, off course, viewed from above. A pulse from A meeting the starboard stern around 1 second before meeting the starboard bow would produce a yaw-and-back. A pulse from the B-direction produces a similar yaw-and-back.

Such pitch cues might indicate one was on a *dilep*. What about cues for heading in the wrong direction so as to lose track? Figure 14 illustrates a boat heading several degrees to the right of the intended course. If we (inaccurately and simplistically — see below) consider an individual wave from A as a pulse meeting first the starboard stern, then 1 second later (on the numbers used here) the starboard bow, it might provide a temporary clockwise yaw that is reversed after a second. The swell from B produces the identical type of yaw-and-back. But the yaw-and-backs from A and B will only be in sync in the region of minimum pitch; in the region of maximum pitch they will be out of sync, i.e. some 4 seconds apart. Regardless of such synchronisation, the 'handedness' of each such yaw-and-back is a clue as to which way the boat is deviating off course; here the initial yaw-right, closely followed by a return back, indicates that the boat is currently aimed to the right of the intended course.

We should stress again that it is inaccurate and simplistic to treat such waves as pulses — they are up-and-down motion more than lateral motion — so this is no more than a first essay at identifying possible cues. The more complex real effects may here be combined pitch-and-roll more than yaw, and will be in part dependent on e.g. the natural roll period of the boat.

Though these are crude simplifications and the real world will be much more subtle, we can nevertheless draw two lessons. Firstly, any such cues will depend in large part on the dimensions of the boat and its pitch, yaw, roll etc. characteristics. The boat is not just a steerable means of motion but also a sensing mechanism in interaction with the waves. Secondly, the cues available to the navigator will likely fluctuate or alternate as the boat moves along the *dilep* between regions of maximum pitch, here 50m apart; around once a minute for a boat travelling at 2 knots.

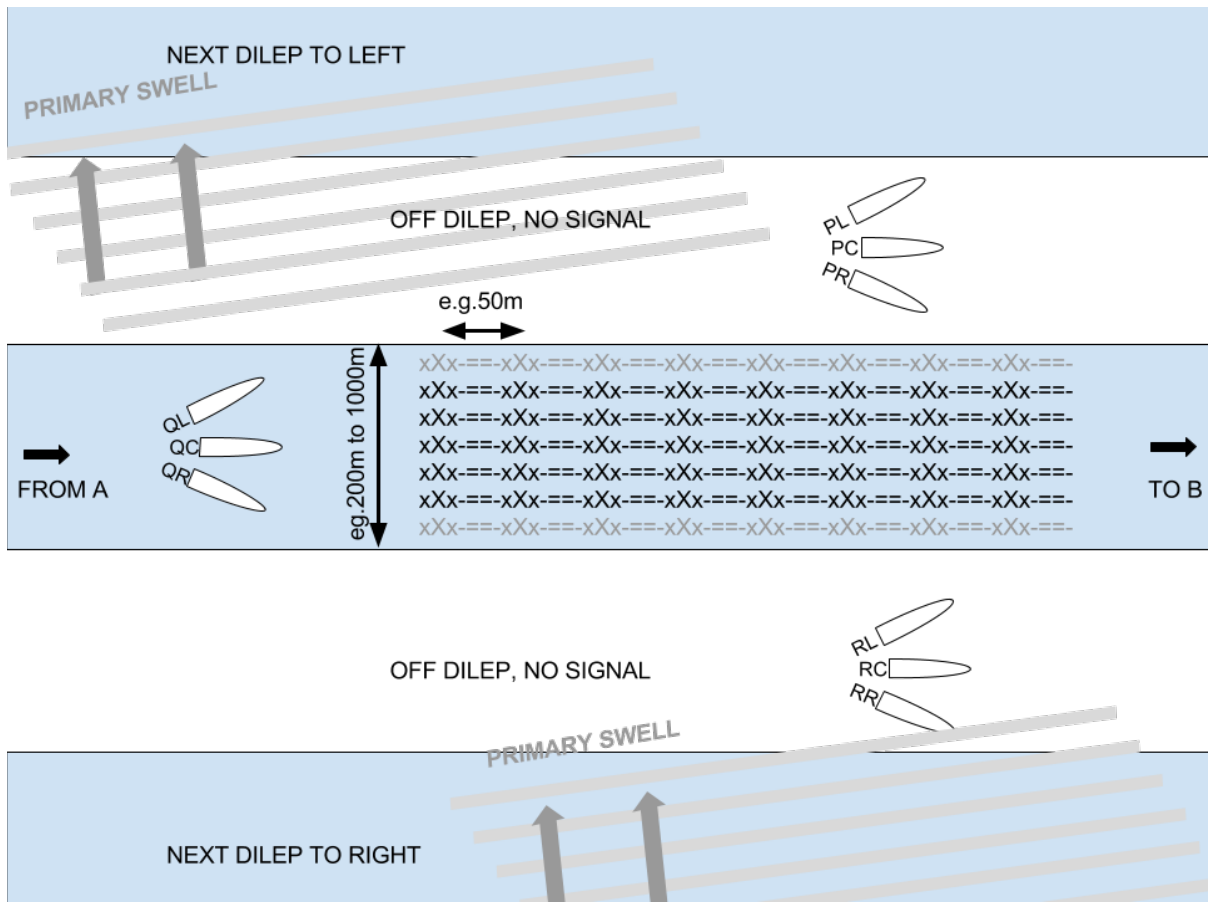

Figure 15. A schematic of features around one *dilep* in our model, to be compared with that of antecedent models in Figure 1. The primary swell is only partly shown. Along the main *dilep* shown, regions of maximum pitch (xXx) alternate with regions of maximum heave (-==-), and both these fade in prominence towards the edges of the *dilep*. Boats QL, QC and QR are expected to experience distinctive signals corresponding to their headings; the other off-*dilep* boats shown will experience no such signals.

## **Our new model schematised**

Figure 15 gives a schematic highlighting major differences between our model and the antecedent model of Figure 1. Along the central *dilep* shown — and indeed any of the other (near-)parallel *dileps* —- we expect regions of maximal pitch (marked xXx) to occur at 50m intervals (given our assumptions) and we expect the useful signals indicating that a boat's heading is left-of-course, on-course, or right-of course to be most discernible at these peaks of pitch-activation. The intermediate peaks of heave-activation (marked -==-) are likely to be less discernible against the larger pitch movements from the primary swell.

**Testable predictions**

Our model is speculative, based primarily on very idealised conditions, and largely slides over the practical questions of how such tiny signals could be detected from the noise. Our argument has been that, regardless of the sensing difficulties, the discussed invariant patterns appear to be the only ones available to be exploited. The model does provide a number of testable predictions. We can pin exact numbers or ranges onto some of these predictions in idealised scenarios, to be tested analytically or through simulation; other predictions take account of noise and intermittent messy changes in real seas, and so may only be verifiable by voyaging data from the ocean.

1. The primary departure of this model is that rather than one just central *dilep*, we propose numerous parallel ones. We argue that this is compatible with navigators assuming in practice there is only one that they are following. We note that if there was only a single one, then a navigator who had lost track of it and was uncertain whether it was left or right, they would only have a 50% chance of guessing correctly; our multiple parallel *dileps* make that 100%. We predict that such multiple parallel tracks might be observed from satellite photos, or from traversing by boat at right-angles to an inter-island track with instruments or an experienced navigator capable of discerning them.

2. Our model predicts that a navigator currently on a *dilep* with discernible heading signals will be able, just by shifting heading in the same location, to rapidly change between off-left, on-course and off-right signals. This would be distinctly different from the prediction we understand antecedent models to provide, where such signals refer to different track-relative locations and so would require a traverse of the full width of the *dilep* to see all the changes.

3. We quantify the expected distance between two central *dileps*, under steady idealised conditions, as somewhere around 0.7 to 1.0 sqrt(N) wavelengths, where the inter-island distance is N wavelengths. We can likewise predict the decreasing sizes of gaps as one moves further away from the direct AB track.

4. If the A-swell is systematically shorter than the B-swell, we predict the *dilep* patterns will shift towards those pictured in Figure 6. We can quantify the distance between *dilep*-ends: if the distance AB is *n* more A-wavelengths than it is measured in B-wavelengths, the *dilep*-ends divide AB into *n* segments, as shown in Figure 8.

5. We offer 3 possible candidates for the *booj* concept, each predicting a quantifiable scale. The shortest scale is very precise, the distance apart of activation peaks (e.g. peak pitching) of the standing wave of the *dilep*, exactly half the underlying swell wavelength (e.g. here 50m).

6. The medium scale *booj* candidate is based on *dileps* being intermittent: to peter out and be replaced by other *dilep*-segments, displaced nearby but still parallel. The statistics of such underlying changes in conditions, such as different wavetrains succeeding each other, would then translate into a quantifiable prediction for the typical lengths of *dilep*-segments.

7. The longest scale *booj* candidate relies on conditions as in (4) above, which clearly defines a quantifiable scale.

8. We discussed the Marshallese school of thought that treats *dileps* as arising from opposing swells crossing the inter-island track from each side. Though our interpretation suggests such swells are conceptual tools rather than real, it also tentatively suggests a testable prediction: ask the users of such a conceptual scheme to estimate the wavelength and period of such notional opposing swells. Distance cues, if discernible, should indicate the wavelength as much longer than primary swell wavelength, whereas timing cues should indicate the same period (breaking the usual wavelength/period relationship).

**Conclusion**

Occam's razor suggested the brutally minimal model used. If we made the wrong kind of simplification, all our speculations come to naught. This model provides just one candidate class of invariant, and our conclusions follow inevitably. Our idealised model does lead to testable predictions, several with hard numbers attached, to be checked against analysis or simulations. Our discussion of a messier, less idealised real ocean — less synchronised, perhaps intermittent wavetrains, from irregular reflecting islands — suggests that some such predictions will survive. These could be tested against satellite photos, experiments by small boat navigators in the right conditions, and ethnographic reports based on traditional navigational lore. But we are well aware that we have largely ignored almost all of the practical issues of how to detect such weak signals amidst all that noise.

If traditional navigators of the Pacific have indeed detected and used such *dilep* signals in the way we propose here, this raises the possibility that modern smartphone technology, with inbuilt gyroscopic sensors and accelerometers, might be exploited to similar ends. The underlying mathematics here is much the simpler part of the problem; the achievements of generations of Pacific sailors with no tools beyond the boat itself, accumulating and passing on such expertise — requiring underlying theories (however conceived) *and* practical sensing techniques *and* developed intuitions — are awe-inspiring on an altogether different scale. It is important to preserve and re-create such skills and knowledge.